# Three-dimensionally Enlarged Photoelectrodes by a Protogenetic Inclusion of Vertically Aligned Carbon Nanotubes into CH$_3$NH$_3$PbBr$_3$ Single Crystals

*Pavao Andričević\*[1], Márton Kollár[1], Xavier Mettan[1], Bálint Náfrádi[1], Andrzej Sienkiewicz[1,2], Dóra Fejes[1], Klára Hernádi[3], László Forró[1], Endre Horváth[1]*

1Laboratory of Physics of Complex Matter (LPMC), Ecole Polytechnique Fédérale de Lausanne, Centre Est, Station 3, CH-1015 Lausanne, Switzerland;
2ADSresonances SARL, Route de Genève 60B, CH-1028 Préverenges, Switzerland;
3Department of Applied and Environmental Chemistry, University of Szeged, Rerrich Béla tér 1, HU-6720 Szeged, Hungary

\*pavao.andricevic@epfl.ch

## ABSTRACT

We demonstrate that single crystals of methylammonium lead bromide (MAPbBr$_3$) could be grown directly on vertically aligned carbon nanotube (VACNT) forests. The fast-growing MAPbBr$_3$ single crystals engulfed the protogenetic inclusions in the form of individual CNTs, thus resulting in a three-dimensionally enlarged photosensitive interface. Photodetector devices were obtained, detecting low light intensities (~20 nW) from UV range to 550 nm. Moreover, a photocurrent was recorded at zero external bias voltage which points to the plausible formation of a p-n junction resulting from interpenetration of MAPbBr$_3$ single crystals into the VACNT forest. This reveals that vertically aligned CNTs can be used as electrodes in operationally stable perovskite-based optoelectronic devices and can serve as a versatile platform for future selective electrode development.

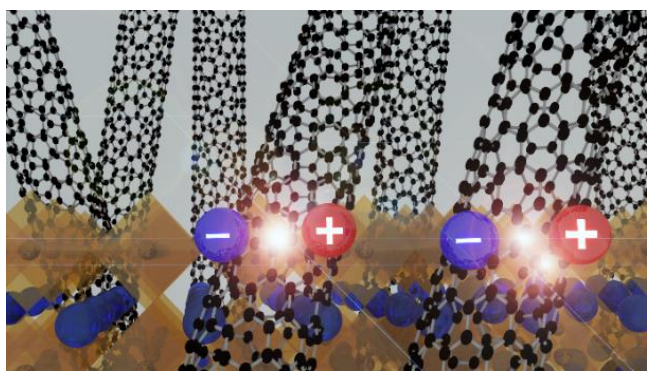

# INTRODUCTION

Semiconducting organic-inorganic lead halide perovskites,[1-4] like methylammonium lead triiodide, $CH_3NH_3PbI_3$ ($MAPbI_3$), or methylammonium lead tribromide $CH_3NH_3PbBr_3$ ($MAPbBr_3$), hold great promise in becoming the most potent game-changer in the photovoltaic industry during the next decade. Recently, these materials unraveled outstanding optoelectronic properties, such as high absorption coefficient, high carrier mobility, long and balanced carrier electron and hole transport and low excitation binding energy.[5-7] Small area perovskite solar cells, with the cell active area of ~0.1 $cm^2$, have recently reached a certified efficiency of 22.1%[8], whereas the devices having larger sizes achieved efficiencies of 20[9] and 10 %[10], for 1 and 70 $cm^2$ size devices, respectively.

Next to the already reported efficient solar cells, fast photon sensing, even at low illumination intensities,[11] electroluminescence,[12] record efficiency in lasing,[13] gas sensing,[14] relatively high thermoelectric figure of merit,[6] memristive effects,[15] and most recently optically switched ferromagnetic behavior in magnetic ions doped $MAPbI_3$ single crystals[16] has also been demonstrated. All these appealing physicochemical properties continuously render organic-inorganic lead halide perovskite attractive and economically viable alternatives for conventional semiconductors.

Nevertheless, to make metal halide perovskite solar cells commercially competitive, there are still several challenges to overcome. Unlike the perovskite absorber layer itself, which is most commonly deposited by a simple and cheap solution process method, the metal electrodes of the state-of-the-art perovskite solar cell are noble metal based (Au/Ag) and deposited by high vacuum thermal evaporation, hence relatively expensive.[17,18] This might be one of the obstacles hindering large-scale commercialization and requires searching for low-cost and abundant materials as replacement of noble metal electrodes. Besides the expectable costs, at present, the long term device stability is one of the most critical issues, as it is yet unclear which specific mechanisms lead to performance loss and subsequent device failure.[19] Even noble metal gold has been reported to migrate inside the cell at working condition temperatures (70 °C), thus severely affecting the device performance metric,[20] while silver and aluminum will oxidize in the long run.[18] Similarly to the metal electrodes, the layers of hole-transport materials (HTM) of today's best performing perovskite solar cells also represent cost and device stability concerns.[19]

In the search for viable alternatives, carbon materials are considered both as potential HTM[17,18,21,22] and counter electrode materials. Fullerenes[23] and graphene[24] oxide were first introduced as substitutes for hole conductors in heterojunction perovskite solar cells, thus facilitating to achieve efficiencies up to 12%. Follow up on, single wall (SWCNTs) and multi wall carbon nanotubes (MWCNTs) have been integrated in various other optoelectronic applications due to their direct band gap (SWCNTs)[21] and outstanding electronic and mechanical properties.[25] The highest efficiencies (18.8%) of carbon-containing perovskite solar cells were fabricated using double-layer structure of polymer-wrapped SWCNTs and undoped spiro-OMeTAD as the hole-transporting p-type layer.[26]

SWCNTs are especially interesting since they have been observed to conduct electrons ballistically. Thus, there is no carrier scattering or energy dissipation in the body of these one-dimensional conductors. This unique carrier-transport regime endows SWCNTs with very high charge-carrier mobilities, being of ~1000 times larger than in bulk silicone.[27] Spina *et al.* fabricated perovskite-based photo-field-effect transistors (FETs) using carbon nanotube and graphene channels, achieving ultrahigh responsivity as of $2.6 \times 10^6$ A/W and demonstrating charge transfer.[11,28,29] According to these findings, CNT-based electrodes show comparable performance to today's most widely used architectures. More importantly, carbon nanomaterials provided additional benefits such as stability[22], reduction of the hysteretic and drift effects,[23,30] flexibility[31] and semitransparency.[17]

A special set of reaction conditions allow carbon nanotubes to self-assemble into vertically oriented cellular arrays during growth on a substrate. It has been demonstrated that this special architecture of CNTs, *i.e.* of a vertically aligned carbon nanotube (VACNT) forest, can grow extremely fast (millimeter-scale height in a 10 min. growth time) and therefore its preparation protocol has been named the supergrowth (SG) method.[32] In fact, the SG method has recently come to the focus of research in nanoscale sciences because it enables rapid synthesis of high-purity CNTs with high degree of order and aspect ratio, thus allowing for good controllability and easy manipulation.

The most efficient way to produce VACNTs is by catalytic chemical vapor deposition (CVD). Transition metals, such as Fe, Co, Ni or their alloys are commonly used as catalyst, while $Al_2O_3$, MgO and $SiO_2$ are the predominant oxide supports.[33-37] It has been shown that aligned CNTs have a large number of short transfer paths for electrons and ions and therefore exhibit a decrease in charge transfer resistance compared to their non-aligned counterparts.[38] Moreover, electrical conductivity values registered in parallel to the nanotubes are 50 times higher compared to those obtained in the perpendicular direction.[39] These traits may be appealing in applications for energy conversion and storage devices. For instance, Li *et al.* reported that VACNTs can be efficiently used as counter electrodes in dye-sensitized solar cells, exhibiting a superior fill factor, short-circuit photo-current density and open-circuit voltage when compared to tangled CNT electrodes.[38]

CNT forests are soft, flexible, spongelike aerogel materials. Nanoindentation studies revealed excellent mechanical properties of the CNT forest, such as high strength and stiffness, while also demonstrating an elastic nature at low indentation forces.[40] As a result, CNTs have the potential for application as dry contact electrodes appropriate to combine with relatively rough surfaces.

Currently, in the perovskite-based photovoltaic devices, semiconducting photoactive perovskite layers are mostly deposited as thin polycrystalline films. Such thin films usually contain a large density of charge traps at the grain boundaries. Since single crystals are considered to have a reduced number of grain boundaries, it is expected that single-crystal-based devices will possess enhanced optoelectronic properties.

Therefore, the purpose of this study was to integrate VACNTs into perovskite-based photodetector devices. To this end, we combined highly efficient light absorbers in the form of single crystal $MAPbBr_3$ (where MA = $CH_3NH_3^+$) with a forest of vertically aligned carbon nanotubes (VACNT). In particular, we observed an unprecedented growth of $MAPbBr_3$ crystals interpenetrating the VACNT forest, thus leading to the formation of large area three-dimensional (3D) perovskite-carbon interfaces. Hence, the structure thus obtained consisted of protogenetic inclusion in the form of vertically aligned CNTs that were engulfed by the fast-growing $MAPbBr_3$ single crystals. Moreover, the observed generation of photocurrent at this intimate $MAPbBr_3$ – CNTs interface, which occurred without external bias, suggests an *in situ* formation of a p-n junction during the crystal inclusion process.

**EXPERIMENTAL SECTION**

*Crystal growth.* Crystals of the methylammonium lead tribromide were synthesized by solution growth. The 3.3 mmol lead (II) acetate trihidrate ($Pb(ac)_2 \times 3H_2O$, >99.9%) was reacted with 6 ml saturated HBr solution (48 wt % HBr in $H_2O$). The formed $PbBr_2$ precipitate is stable in the acidic solution. The respective amount (3.30 mmol) of methylamine ($CH_3NH_2$) solution (40 wt % in $H_2O$) was pipetted into the 5 °C ice cooled solution of $PbBr_2$. The cold solution avoids the evaporation of methylamine during the exothermic reaction. Orange colored microcrystallites of $CH_3NH_3PbBr_3$ were formed. The MAPbBr$_3$ crystals were recrystallized in a temperature gradient of 15 °C in the acidic media to get transparent, high purity crystals.

The $CH_3NH_3PbCl_3$ crystals were synthesised by solution growth, using the same molar amounts and receipt as for the MAPbBr$_3$ synthesis, but using HCl (36 wt % in $H_2O$) reagent instead of the HBr solution.

MAPbBr$_3$ single crystals were grown by inverse temperature crystallization from its saturated solution in DMF. 0.8 g MAPbBr$_3$ was dissolved per cm$^3$ of DMF at room temperature. The substrate was immersed in the solution, and a MAPbBr$_3$ seed crystal was placed on the top of the VACNT. Crystal growth was initialized by increasing the temperature of the solution from room temperature to 40 °C with a heating rate of 5 °C/h. We observed that the fast-growing seed single crystals gradually protruded and engulfed the individual nanotubes. These type of inclusions, when the original form of the included mineral is preserved in the host crystal is categorized as a protogenetic inclusion in mineralogy. The slow heating rate suppress the formation of new seed crystals in the supersaturated solution and our MAPbBr$_3$ seed crystal can grow rapidly on the substrate. When the required size of crystal was reached, the MAPbBr$_3$ – VACNT composite was removed from the solution, wiped and dried.

MAPbCl$_3$ single crystals were grown from room temperature saturated solution of MAPbCl$_3$ in 1:1 DMF : DMSO solvent mixture, and were grown by inverse temperature crystallization method.

*Optoelectronic characterization.* All measurements of performances of the devices were done in ambient conditions at room temperature. The junction characteristics have been determined by two point resistivity measurements, using golden wires contacted with Dupont 4929 silver epoxy as electrical leads. One of the contacts is positioned directly on the perovskite single crystal, whereas the second one touches the CNT forest only. A Keithley 2400 source meter allowed us to measure the current with < 0.1 nA resolution, while tuning the applied bias voltage, in dark and under visible and UV light illumination. Photocurrent measurements at low light intensities were done by choosing a wavelength of 550 nm, within the spectral response of our device, and with which we are able to achieve high enough intensities of light that can be detected. The photocurrent spectrum was acquired while measuring the current response to the incident light at 2 V bias voltage and by sweeping over wavelengths (2 nm/sec). The wavelengths were set thanks to a monochromator (Horiba Micro HR), while different intensities of light were achieved by closing and opening slits in the light path.

*Photoluminescence measurement.* The photoluminescence (PL) spectra were acquired using a custom-designed setup, which was based on a combination of an inverted biological epi-fluorescent microscope (TC5500, Meiji Techno, Japan) and a spectro-fluorometer (USB 2000+XR, Ocean Optics Inc., USA). The PL signals were recorded upon excitation at $\lambda_{exc}$= 470 nm. This excitation light wavelength was filtered out from the emission of the microscope's Mercury vapor 100-W lamp by implementing a dedicated set of Meiji Techno filters, Model 11001v2 Blue.[41]

**RESULTS AND DISCUSSION**

MAPbBr$_3$ single crystals were grown by inverse temperature crystallization from its saturated N,N-dimethylformamide (DMF) solution.[42-44] The first junctions (mechanical contacts) were made by dry pressing the MAPbBr$_3$ single crystal against the surface of the CNT forests, which resulted in the formation of numerous direct contacts of the crystal with CNTs terminations.

Two point resistivity measurements were performed with tungsten probe needles as electrical leads. The devices (Figure S1) showed diode characteristics, typical for a metal-semiconductor heterojunction. Importantly, higher values of photocurrent were obtained with using higher CNT forests. The reason behind this observation is that, due to the CNTs millimeter-lengths, higher forests accommodate better the actual crystal shape, thus leading to larger contact areas. However, when trying to append a single MAPbBr$_3$ crystal atop the VACNT forest, higher mechanical stress was exerted on the CNTs. Consequently, the CNTs plastically deformed and numerous pinholes were formed within the VACNT forest. Such pinholes made it possible for the MAPbBr$_3$ single crystal to get in direct contact with the silicon substrate, thus resulting in short circuits.

Therefore, a radically different approach had to be explored towards designing an efficient and operationally stable photo-detecting device based on the combination of vertically aligned CNTs and MAPbBr$_3$.

The most important stages of the herein proposed growth of MAPbBr$_3$ single crystals on the formerly prepared VACNT forest are shown in Figure 1. In particular, in Figure 1a, in a series of photographs taken at different time intervals as the progressive volumetric growth of MAPbBr$_3$ crystals on the vertically aligned CNTs is shown. As can be seen, the deposit of the orange-colored material, which is characteristic for MAPbBr$_3$, was increasing steadily as a function of time. In fact, since the occupation volume of the vertically aligned CNTs represents only ~5 vol %,[36] MAPbBr$_3$ single crystals were grown directly on the VACNT forest. The subsequent technological steps corresponding to a direct growth MAPbBr$_3$ single crystals on the VACNT forest are schematically depicted in Figure 1b.

Prior to the process of crystallogenesis, a saturated solution of MAPbBr$_3$ precursors in DMF was prepared at room temperature. Subsequently, the VACNT forest grown on a silicon substrate[33] (Figure 1c) was immersed in the solution. A small MAPbBr$_3$ single crystal (2 × 2 × 1 mm$^3$) was used as a seed crystal and was placed atop of the VACNT forest.

The crystal growth was initialized by warming up the solution from room temperature to 40 °C at a slow rate (5 °C/h). Interestingly, in addition to the crystal growth on the VACNT canopy surface, the fast-growing crystalline fractions also gradually protruded into the CNTs forest, thus interpenetrating and engulfing the individual carbon nanotubes (Figure 1d). The presence of such protogenetic inclusions in the form of vertically aligned CNTs within the MAPbBr$_3$ crystals resulted in a three-dimensionally enlarged photosensitive interface.

In the final technological step, the CNTs were mechanically detached from the silicon substrate leading to a neat MAPbBr$_3$ and CNT junction, as shown in Figure 1e. The inverse temperature crystal growth was performed with the MAPbCl$_3$ compound (Figure S2). Similarly to the MAPbBr3 perovskite, the single crystals of MAPbCl$_3$ grew around the CNT forest engulfing individual CNTs. This indicates that the protogenetic inclusion of carbon nanotubes and other foreign objects might be feasible for the whole family of halide perovskites, which grow by inverse temperature crystallization.

To confirm that CNTs are protogenetic inclusions in the perovskite single crystals, scanning electron microscopy (SEM) images of the interface were acquired. As one can see in Figure 1f,g a thick layer of CNTs, uniform thickness of ~ 40 μm, protrudes above the surface of the MAPbBr$_3$ single crystal. Specifically, under higher magnification, it is clearly visible that individual CNTs are embedded within the crystalline matrix (Figure 1h).

Stability at atmospheric conditions is one of the major concerns of perovskite-based devices. Due to the high sensitivity of MAPbBr$_3$ and other organic-inorganic halide perovskites to water, they tend to hydrolyze in the presence of moisture, which results in degradation of the perovskite.[45] Furthermore, because of the presence of the lead compound, recent studies showed that lead based perovskite could have a negative effect on health and the environment.[46] Therefore, the photodetector designed herein was encapsulated into polydimethylsiloxane (PDMS). PDMS is optically transparent, inert, nontoxic, nonflammable, well-known for its flexibility and chemical resistance.[47]

In order to characterize the photo induced charge transfer phenomena of the MAPbBr$_3$–CNT junction, two point electrical measurements (current vs. voltage) were performed under visible and UV light illumination. A schematic presentation of the interface and its charge transfer phenomena is illustrated in Figure 2a. The spectral responsivity, *i.e.* the photocurrent and photoluminescence (PL) spectra of the MAPbBr$_3$ single crystal-based photodetector encapsulated in PDMS are shown in Figure 2b. A characteristic drop of spectral sensitivity at around 540 nm is observed, which corresponds to the 2.3 eV band gap of MAPbBr$_3$.[42] As shown in Figure 2b, the observed drop in the photocurrent coincides spectrally with the peak of green PL emission, which is also centered at 540 nm. This PL peak corresponds to the near-edge band-gap emission of MAPbBr$_3$ and is consistent with previous literature reports.[48] PDMS does not have any influence on the spectral responsivity because the results do not differ from those obtained before encapsulation (Supporting Information, Figure S3).

*I-V* photocurrent characteristics were obtained under illumination of a white fluorescent light source with intensity of 1.02 mW/cm$^2$. As can be seen in Figure 2c, the MAPbBr$_3$/VACNT composite samples reveal diode-like behavior, with a forward onset at ~1.0 V and an almost complete saturation of current in the reverse bias direction. The dark current follows the same behavior. Furthermore, sweeping the applied voltage in a fixed range but at different rates leads to different currents. This is the result of electrical poling of the perovskite crystal: when a bias voltage of 1.5-2.0 V is applied, the current increases before saturating. The slow response suggests that ionic migration under an applied electric field (poling mechanism) is likely the reason of such behavior.[49] The same mechanism can also explain the hysteresis effect (Figure 2c). Besides the observed photocurrent at higher bias voltages (>1 V), when analyzing in detail the region around 0 V bias, a finite photocurrent was detected, even when no voltage was applied. In the inset of Figure 2d, one can observe that this photocurrent, $I_{ph}$, is of 23 nA at zero bias, thus pointing to the possibility of the formation of a p-n junction. Photocurrent measurements performed before encapsulation revealed similar results (Figure S4).

Measurements of the photocurrent at ambient conditions were done under low light intensities to determine the responsivity of the light sensing device. On-off measurements were performed for light intensities ranging from 18 to 250 nW. The corresponding on-off characteristics of the transient photoresponse, collected at bias voltages of 2 and 0 V, are shown in parts a and b, respectively, of Figure 3. From these results the responsivity $R$ of the device can be calculated (Figure 3c). $R$ is defined as the ratio of the photocurrent $I_{ph}$ and the intensity of light $P_{light}$.

$$R = \frac{I_{light} - I_{dark}}{P_{light}} = \frac{I_{ph}}{P_{light}}$$

The responsivity $R$ of the device is of the order of magnitude of $10^{-1}$ and $10^{-3}$ A/W, at 2 and 0 V bias, respectively. For both bias voltages the responsivity increases rapidly with lower intensities of light (Figure 3c). On-off measurements at 0 V bias, shown in Figure 3b, were performed before encapsulation of the detecting device. Other measurements performed for various photo-detecting structures before encapsulation revealed the same characteristics with different absolute values of the photocurrent depending on the geometry and quality of the sample. These latter results are shown in Figure S5.

Upon applying the bias voltage of 2 V, under illumination, the photocurrent increased stepwise, thus corresponding to a fast regime (Figure 3a,d). Moreover, the photocurrent continued to increase for longer times of illumination until it saturated at a finite value. In contrast, such behavior was not observed for the bias voltage of 0 V. We infer that the ion migration within the perovskite crystal (poling) could be responsible for this type of the photocurrent response.

Apart from the responsivity, other important performance benchmark of a photodetector are the photoresponse times, *i.e.* the rise time, $t_{rise}$ and the fall time, $t_{fall}$. The former we defined as the time elapsed for the rise of the current intensity from 10 to 90% of the peak output current amplitude ("On-state"), with respect to the initial current ("Off-state"), while the latter is the time elapsed from 90 to 10% of the maximum current (Figure 3d). Our device exhibits a rise time of 0.4 s, while the fall time is 0.53 s. To put these numbers into perspective a comparison to state-of-the-art photodetectors is shown in Table S1. The MAPbBr3/VACNT photodetectors have achieved technological progress in terms of the operational voltage (low bias or zero-bias mode) as well as lowering the detection limit to 18 nW light power.

Finally, we studied the *I-V* characteristics of the junction at low temperatures under light (Figure 3e) and in dark (Figure S6). The photodetector device showed the same diode characteristics at all temperatures down to 30 K. Moreover, we observed that under illumination with visible light (intensity 0.77 mW/cm$^2$) the photocurrent at a bias voltage of 2 V increased at lower temperatures. Furthermore, the hysteresis of I-V curves, depicted in Figure 3e (inset), disappears upon lowering the temperature, supporting the presence of mobile ions. Indeed, the kinetics of ion migration is slowed down at lower temperatures. As shown in Figure 3f, the current was also measured under light at a bias voltage of 2 V, while continuously sweeping the temperature from 300 to 20 K and back. The standard transition from the orthorhombic to the tetragonal phase was observed around 148 K[50] as well as a residual resistivity increase at lower temperatures.

Carbon nanotube forests are a mixture of single-, double-, and triple-wall CNTs[37] that are either true metals or a semiconductors. Multiwall CNTs have been reported to reveal metallic behavior, thus acting as ballistic conductors.[51] Therefore, they can form Schottky or metal-semiconductor junctions with semiconducting perovskite single crystals. The Schottky barrier formed at the interface allows the flow of thermally excited electrons over the barrier under forward bias, while preventing flow under reverse bias. This explains why the device designed herein revealed diode-like *I-V* behavior and worked as a photodetector even if no voltage was applied.

Interestingly, this behavior of a photo-current generation without application of external voltage, was not observed for the devices prepared using the "dry contact" method. Therefore, we speculate that during the *in situ* growth, the carbon nanotubes were functionalized, thus becoming p-type semiconductors. Notwithstanding, we cannot exclude the formation of chemical bonds at the junction interface, either. Moreover, bromide, as well as other halides, are capable to chemically dope single-wall and multi-wall CNTs.[52-55] Hence, the photodetector device could function as a p-n junction, as shown in Figure S7. This might provide a significant driving force for the hole injection from the perovskite layer to the CNT forest, thus collecting and transporting holes to the external circuit.[56] Furthermore, recently a vast number of publications present the use of SWCNTs or MWCNTs as p-type contacts in perovskite devices without relying on functionalization.[17,18,21,22,57,58] Adding an electron-selective contact as a back electrode could be advantageous to eliminate the hole transfer at the back electrode, thus improving the performance of the existing device.

**CONCLUSIONS**

In summary we report, for the first time, the growth of MAPbBr$_3$ single crystals interpenetrating VACNT forests, engulfing individual nanotubes as protogenetic inclusions. This approach resulted in the formation of a three-dimensionally enlarged photosensitive interface. Millimeter-sized photodetector devices were obtained, capable of detecting low light intensities (20-200 nW) from UV range to 550 nm. Importantly, photocurrent was measured at zero external bias voltage, which points to the formation of a p-n junction during the single crystal inclusion into the vertically aligned carbon nanotube forest. Therefore, pristine or functionalized VACNTs are potential candidates for fabrication of metallic or selective semiconducting, p or n-type, electrodes, thus leading towards future operationally stable perovskite-based optoelectronic devices.

**ACKNOWLEDGMENT**

This work was supported by the Swiss National Science Foundation (No. 160169) and the ERC advanced grant "PICOPROP" (Grant No. 670918). The authors are also very grateful to the financial support provided by the OTKA NN114463. K.H. acknowledges the financial support of GINOP-2.3.2- 15-2016- 00013 project.

**REFERENCES**

(1)     Lee, M. M.; Teuscher, J.; Miyasaka, T.; Murakami, T. N.; Snaith, H. J. Efficient Hybrid Solar Cells Based on Meso-Superstructured Organometal Halide Perovskites. *Science* 2012, *338*, 643−647.

(2) Kim, H.-S.; Lee, C.-R.; Im, J.-H.; Lee, K.-B.; Moehl, T.; Marchioro, A.; Moon, S.-J.; Humphry-Baker, R.; Yum, J.-H.; Moser, J. E.; *et al.* Lead Iodide Perovskite Sensitized All-Solid-State Submicron Thin Film Mesoscopic Solar Cell with Efficiency Exceeding 9%. *Sci. Rep.* 2012, *2*, 591.

(3) Liu, M.; Johnston, M. B.; Snaith, H. J. Efficient Planar Heterojunction Perovskite Solar Cells by Vapour Deposition. *Nature* 20139, *501*, 395–398.

(4) Burschka, J.; Pellet, N.; Moon, S.-J.; Humphry-Baker, R.; Gao, P.; Nazeeruddin, M. K.; Grätzel, M. Sequential Deposition as a Route to High-Performance Perovskite-Sensitized Solar Cells. *Nature* 2013, *499*, 316–319.

(5) Xing, G.; Mathews, N.; Lim, S. S.; Lam, Y. M.; Mhaisalkar, S.; Sum, T. C. Long-Range Balanced Electron-and Hole-Transport Lengths in Organic-Inorganic CH3NH3PbI3. *Science* 2013, *342*, 344-347.

(6) Mettan, X.; Pisoni, R.; Matus, P.; Pisoni, A.; Jacimovic, J.; Náfrádi, B.; Spina, M.; Pavuna, D.; Forró, L.; Horváth, E. Tuning of the Thermoelectric Figure of Merit of CH3NH3MI3 (M=Pb,Sn) Photovoltaic Perovskites. *J. Phys. Chem. C* 2015, *119*, 11506-11510.

(7) Pisoni, A.; Jaćimović, J.; Barišić, O. S.; Spina, M.; Gaál, R.; Forró, L.; Horváth, E. Ultra-Low Thermal Conductivity in Organic–Inorganic Hybrid Perovskite CH3NH3PbI3. *J. Phys. Chem. Lett.* 2014, *5*, 2488–2492.

(8) NREL Efficiency Chart. http://www.nrel.gov/ncpv/images/efficiency_chart.jpg (accessed November 3, 2016)

(9) Li, X.; Bi, D.; Yi, C.; Decoppet, J.-D.; Luo, J.; Zakeeruddin, S. M.; Hagfeldt, A.; Gratzel, M. A Vacuum Flash-Assisted Solution Process for High-Efficiency Large-Area Perovskite Solar Cells. *Science.* 2016, *353*, 58–62.

(10) Priyadarshi, A.; Haur, L. J.; Murray, P.; Fu, D.; Kulkarni, S.; Xing, G.; Sum, T. C.; Mathews, N.; Mhaisalkar, S. G.; Heo, J. H.; *et al.* A Large Area (70 Cm$^2$) Monolithic Perovskite Solar Module with a High Efficiency and Stability. *Energy Environ. Sci.* 2016, *9*, 3687-3692.

(11) Spina, M.; Lehmann, M.; Náfrádi, B.; Bernard, L.; Bonvin, E.; Gaál, R.; Magrez, A.; Forrõ, L.; Horváth, E. Microengineered CH3NH3Pb3 Nanowire/Graphene Phototransistor for Low-Intensity Light Detection at Room Temperature. *Small* 2015, *11*, 4824–4828.

(12) Tan, Z.-K.; Moghaddam, R. S.; Lai, M. L.; Docampo, P.; Higler, R.; Deschler, F.; Price, M.; Sadhanala, A.; Pazos, L. M.; Credgington, D.; *et al.* Bright Light-Emitting Diodes Based on Organometal Halide Perovskite. *Nat. Nanotechnol.* 2014, *9*, 687-692.

(13) Deschler, F.; Price, M.; Pathak, S.; Klintberg, L. E.; Jarausch, D. D.; Higler, R.; Hüttner, S.; Leijtens, T.; Stranks, S. D.; Snaith, H. J.; *et al.* High Photoluminescence Efficiency and Optically Pumped Lasing in Solution-Processed Mixed Halide Perovskite Semiconductors. *J. Phys. Chem. Lett.* 2014, *5*, 1421–1426.

(14)     Bao, C.; Yang, J.; Zhu, W.; Zhou, X.; Gao, H.; Li, F.; Fu, G.; Yu, T.; Zou, Z. A Resistance Change Effect in Perovskite CH3NH3PbI3 Films Induced by Ammonia. *Chem. Commun.* 2015, *51*, 15426–15429.

(15)     Xiao, Z.; Yuan, Y.; Shao, Y.; Wang, Q.; Dong, Q.; Bi, C.; Sharma, P.; Gruverman, A.; Huang, J. Giant Switchable Photovoltaic Effect in Organometal Trihalide Perovskite Devices. *Nat. Mater.* 2014, *14*, 193–198.

(16)     Náfrádi, B.; Szirmai, P.; Spina, M.; Lee, H.; Yazyev, O. V; Arakcheeva, A.; Chernyshov, D.; Gibert, M.; Forró, L.; Horváth, E. Optically Switched Magnetism in Photovoltaic Perovskite CH3NH3(Mn:Pb)I3. *Nat. Commun.* 2016, *7*, 13406.

(17)     Li, Z.; Kulkarni, S. a; Boix, P. P.; Shi, E.; Cao, A.; Fu, K.; Batabyal, S. K. Laminated Carbon Nanotube Networks for Metal Electrode-Free. *ACS Nano* 2014, *8*, 6797–6804.

(18)     Aitola, K.; Sveinbjornsson, K.; Correa Baena, J. P.; Kaskela, A.; Abate, A.; Tian, Y.; Johansson, E. M. J.; Grätzel, M.; Kauppinen, E.; Hagfeldt, A.; *et al.* Carbon Nanotube-Based Hybrid Hole-Transporting Material and Selective Contact for High Efficiency Perovskite Solar Cells. *Energy Environ. Sci.* 2015, *9*, 461–466.

(19)     Wei, D.; Wang, T.; Ji, J.; Li, M.; Cui, P.; Li, Y.; Li, G.; Mbengue, J. M.; Song, D. Photo-Induced Degradation of Lead Halide Perovskite Solar Cells Caused by the Hole Transport Layer/metal Electrode Interface. *J. Mater. Chem. A* 2016, *4*, 1991–1998.

(20)     Domanski, K.; Correa-Baena, J. P.; Mine, N.; Nazeeruddin, M. K.; Abate, A.; Saliba, M.; Tress, W.; Hagfeldt, A.; Grätzel, M. Not All That Glitters Is Gold: Metal-Migration-Induced Degradation in Perovskite Solar Cells. *ACS Nano* 2016, *10*, 6306–6314.

(21)     Habisreutinger, S. N.; Leijtens, T.; Eperon, G. E.; Stranks, S. D.; Nicholas, R. J.; Snaith, H. J. Carbon Nanotube/Polymer Composite as a Highly Stable Charge Collection Layer in Perovskite Solar Cells. *Nano Lett.* 2014, *14*, 5561-5568

(22)     Wang, F.; Endo, M.; Mouri, S.; Miyauchi, Y.; Ohno, Y.; Wakamiya, A.; Murata, Y.; Matsuda, K. Highly Stable Perovskite Solar Cells with All-Carbon Hole Transport Layer. *Nanoscale* 2016, 11882–11888.

(23)     Jeng, J. Y.; Chiang, Y. F.; Lee, M. H.; Peng, S. R.; Guo, T. F.; Chen, P.; Wen, T. C. CH3NH3PbI3 Perovskite/fullerene Planar-Heterojunction Hybrid Solar Cells. *Adv. Mater.* 2013, *25*, 3727–3732.

(24)     Wu, Z.; Bai, S.; Xiang, J.; Yuan, Z.; Yang, Y.; Cui, W.; Gao, X.; Liu, Z.; Jin, Y.; Sun, B. Efficient Planar Heterojunction Perovskite Solar Cells Employing Graphene Oxide as Hole Conductor. *Nanoscale* 2014, *6*, 10505-10510

(25)     Yang, L.; Wang, S.; Zeng, Q.; Zhang, Z.; Peng, L. M. Carbon Nanotube Photoelectronic and Photovoltaic Devices and Their Applications in Infrared Detection. *Small* 2013, *9*, 1225–1236.

(26)     Habisreutinger, S. N.; Wenger, B.; Snaith, H. J.; Nicholas, R. J. Dopant-Free Planar N-I-P Perovskite Solar Cells with Steady-State Efficiencies Exceeding 18%. *ACS Energy Lett.* 2017, *2(3)*, 622-628.

(27) Avouris, P.; Chen, Z.; Perebeinos, V. Carbon-Based Electronics. *Nat. Nanotechnol.* 2007, *2*, 605–615.

(28) Spina, M.; Náfrádi, B.; Tohati, H.; Kamaras, K.; Bonvin, E.; Gaál, R.; Forró, L.; Horváth, E. Ultrasensitive 1D Field-Effect Phototransistor: CH3NH3PbI3 Nanowire Sensitized Individual Carbon Nanotube. *Nanoscale* 2016, *8*, 4888–4893.

(29) Spina, M.; Bonvin, E.; Sienkiewicz, A.; Náfrádi, B.; Forró, L.; Horváth, E. Controlled Growth of CH3NH3PbI3 Nanowires in Arrays of Open Nanofluidic Channels. Sci. Rep. 2016, 6 (1), 19834.

(30) Shao, Y.; Xiao, Z.; Bi, C.; Yuan, Y.; Huang, J. Origin and Elimination of Photocurrent Hysteresis by Fullerene Passivation in CH3NH3PbI3 Planar Heterojunction Solar Cells. *Nat. Commun.* 2014, *5*, 1–7.

(31) Chen, T.; Qiu, L.; Cai, Z.; Gong, F.; Yang, Z.; Wang, Z.; Peng, H. Intertwined Aligned Carbon Nanotube Fiber Based Dye-Sensitized Solar Cells. *Nano Lett.* 2012, *12*, 2568–2572.

(32) Hata, K.; Futaba, D. N.; Mizuno, K.; Namai, T.; Yumura, M.; Iijima, S. Water-Assisted Highly Efficient Synthesis of Impurity-Free Single-Walled Carbon Nanotubes. *Science* 2004, *306*, 1362–1364.

(33) Noda, S.; Hasegawa, K.; Sugime, H.; Kakehi, K.; Zhang, Z.; Maruyama, S.; Yamaguchi, Y. Millimeter-Thick Single-Walled Carbon Nanotube Forests: Hidden Role of Catalyst Support. *Japanese J. Appl. Physics, Part 2 Lett.* 2007, *46*.

(34) Mattevi, C.; Wirth, C. T.; Hofmann, S.; Blume, R.; Cantoro, M.; Ducati, C.; Cepek, C.; Knop-Gericke, A.; Milne, S.; Castellarin-Cudia, C.; *et al.* In-Situ X-Ray Photoelectron Spectroscopy Study of Catalyst- Support Interactions and Growth of Carbon Nanotube Forests. *J. Phys. Chem. C* 2008, *112*, 12207–12213.

(35) Sakurai, S.; Nishino, H.; Futaba, D. N.; Yasuda, S.; Maigne, A.; Matsuo, Y.; Nakamura, E.; Yumura, M.; Hata, K.; Sakurai, S.; *et al.* Role of Subsurface Diffusion and Ostwald Ripening in Catalyst Formation for SWNT Forest Growth. *J. Am. Chem. Soc.* 2012, *134*, 2148-2153.

(36) Robertson, J.; Zhong, G.; Esconjauregui, S.; Zhang, C.; Fouquet, M.; Hofmann, S. Chemical Vapor Deposition of Carbon Nanotube Forests. *Phys. Status Solidi Basic Res.* 2012, *249*, 2315–2322.

(37) Magrez, A.; Smajda, R.; Seo, J. W.; Horváth, E.; Ribič, P. R.; Andresen, J. C.; Acquaviva, D.; Olariu, A.; Laurenczy, G.; Forró, L. Striking Influence of the Catalyst Support and Its Acid-Base Properties: New Insight into the Growth Mechanism of Carbon Nanotubes. *ACS Nano* 2011, *5*, 3428–3437.

(38) Li, S.; Luo, Y.; Lv, W.; Yu, W.; Wu, S.; Hou, P.; Yang, Q.; Meng, Q.; Liu, C.; Cheng, H. M. Vertically Aligned Carbon Nanotubes Grown on Graphene Paper as Electrodes in Lithium-Ion Batteries and Dye-Sensitized Solar Cells. *Adv. Energy Mater.* 2011, *1*, 486–490.

(39) Endrődi, B.; Samu, G. F.; Fejes, D.; Németh, Z.; Horváth, E.; Pisoni, A.; Matus, P. K.; Hernádi, K.; Visy, C.; Forró, L.; *et al.* Challenges and Rewards of the Electrosynthesis of Macroscopic Aligned Carbon Nanotube Array/conducting Polymer Hybrid Assemblies. *J. Polym. Sci. Part B Polym. Phys.* 2015, *53*, 1507–1518.

(40)     Qi, H. J.; Teo, K. B. K.; Lau, K. K. S.; Boyce, M. C.; Milne, W. I.; Robertson, J.; Gleason, K. K. Determination of Mechanical Properties of Carbon Nanotubes and Vertically Aligned Carbon Nanotube Forests Using Nanoindentation. *J. Mech. Phys. Solids* 2003, *51*, 2213–2237.

(41)     Mor, F. M.; Sienkiewicz, A.; Magrez, A.; Forró, L.; Jeney, S. Single Potassium Niobate Nano/microsized Particles as Local Mechano-Optical Brownian Probes. *Nanoscale* 2016, *8(12)*, 6810-6819.

(42)     Saidaminov, M. I.; Abdelhady, A. L.; Murali, B.; Alarousu, E.; Burlakov, V. M.; Peng, W.; Dursun, I.; Wang, L.; He, Y.; Maculan, G.; *et al.* High-Quality Bulk Hybrid Perovskite Single Crystals within Minutes by Inverse Temperature Crystallization. *Nat. Commun.* 2015, *6*, 7586.

(43)    Kadro, J. M.; Nonomura, K.; Gachet, D.; Grätzel, M.; Hagfeldt, A. Facile Route to Freestanding CH3NH3PbI3 Crystals Using Inverse Solubility. Sci. Rep. 2015, 5 (2), 11654.

(44)    Maculan, G.; Sheikh, A. D.; Abdelhady, A. L.; Saidaminov, M. I.; Haque, M. A.; Murali, B.; Alarousu, E.; Mohammed, O. F.; Wu, T.; Bakr, O. M. CH3NH3PbCl3 Single Crystals: Inverse Temperature Crystallization and Visible-Blind UV-Photodetector. J. Phys. Chem. Lett. 2015, 6 (19), 3781−3786.

(45)     Niu, G.; Guo, X.; Wang, L. Review of Recent Progress in Chemical Stability of Perovskite Solar Cells. *J. Mater. Chem. A* 2015, *3*, 8970–8980.

(46)     Benmessaoud, I. R.; Mahul-Mellier, A.-L.; Horváth, E.; Maco, B.; Spina, M.; Lashuel, H.; Forró, L. Health Hazard of the Methylammonium Lead Iodide Based Perovskites: Cytotoxicity Studies. *Toxicol. Res.* 2015, *5*, 407-419

(47)     Han, J. M.; Han, J. W.; Chun, J. Y.; Ok, C. H.; Seo, D. S. Novel Encapsulation Method for Flexible Organic Light-Emitting Diodes Using Poly(dimethylsiloxane). *Jpn. J. Appl. Phys.* 2008, *47*, 8986–8988.

(48)     Schmidt, L. C.; Pertegas, A.; Gonzalez-Carrero, S.; Malinkiewicz, O.; Agouram, S.; Espallargas, G. M.; Bolink, H. J.; Galian, R. E.; Perez-Prieto, J. Nontemplate Synthesis of CH 3NH3PbBr3 Perovskite Nanoparticles. *J. Am. Chem. Soc.* 2014, *136*, 850–853.

(48)     Yi, H. T.; Wu, X.; Zhu, X.; Podzorov, V. Intrinsic Charge Transport across Phase Transitions in Hybrid Organo-Inorganic Perovskites. *Adv. Mater.* 2016, 6509–6514.

(50)     Onoda-Yamamuro, N.; Matsuo, T.; Suga, H. Dielectric Study of CH3NH3PbX3 (X = Cl, Br, I). *J. Phys. Chem. Solids* 1992, *53*, 935–939.

(51)     Schönenberger, C.; Forro, L. Multiwall Carbon Nanotubes. *Phys. World* 2000, *13*, 37–41.

(52)     Rao, A. M.; Eklund, P. C.; Bandow, S.; Thess, A.; Smalley, R. E. Evidence for Charge Transfer in Doped Carbon Nanotube Bundles from Raman Scattering. *Nature* 1997, *388*, 257–259.

(53)     Duclaux, L. Review of the Doping of Carbon Nanotubes (Multiwalled and Single-Walled). *Carbon* 2002, *40*, 1751–1764.

(54)     Bulusheva, L. G.; Okotrub, A. V.; Flahaut, E.; Asanov, I. P.; Gevko, P. N.; Koroteev, V. O.; Fedoseeva, Y. V.; Yaya, A.; Ewels, C. P. Bromination of Double-Walled Carbon Nanotubes. *Chem. Mater.* 2012, *24*, 2708–2715.

(55)     Cambedouzou, J.; Sauvajol, J. L.; Rahmani, A.; Flahaut, E.; Peigney, A.; Laurent, C. Raman Spectroscopy of Iodine-Doped Double-Walled Carbon Nanotubes. *Phys. Rev. B* 2004, *69(23)*, 235422.

(56)     Li, Z.; Boix, P. P.; Xing, G.; Fu, K.; Kulkarni, S. A.; Batabyal, S. K.; Xu, W.; Cao, A.; Sum, T. C.; Mathews, N.; *et al.* Carbon Nanotubes as an Efficient Hole Collector for High Voltage Methylammonium Lead Bromide Perovskite Solar Cells. *Nanoscale* 2016, *8*, 6352–6360.

(57)     Wei, Z.; Chen, H.; Yan, K.; Zheng, X.; Yang, S. Hysteresis-Free Multi-Wall Carbon Nanotube-Based Perovskite Solar Cells with a High Fill Factor. *J. Mater. Chem. A* 2015, *0* (C), 1–6.

(58)     Habisreutinger, S. N.; Nicholas, R. J.; Snaith, H. J. Carbon Nanotubes in Perovskite Solar Cells. *Adv. Energy Mater.* 2016, 1–7.

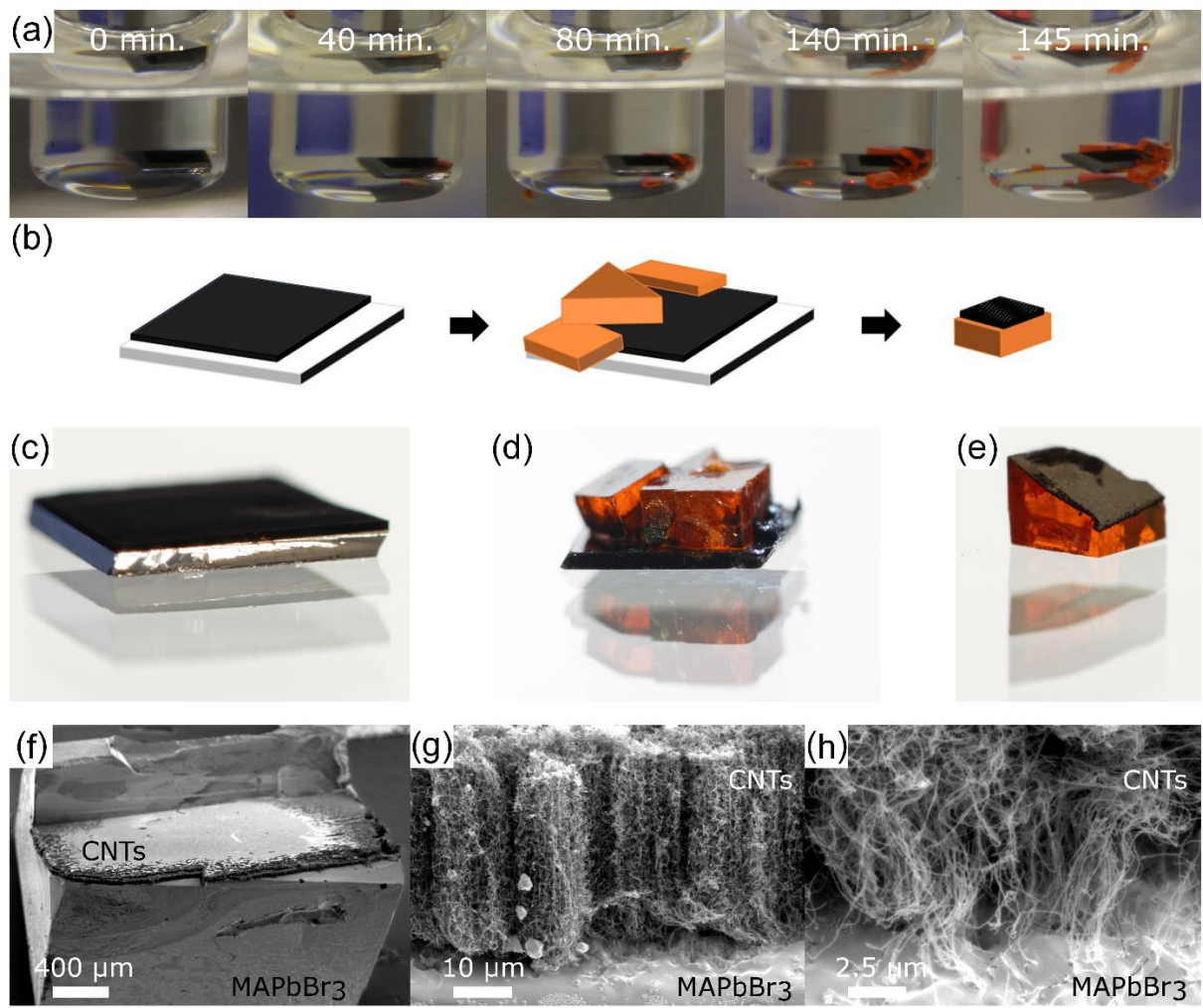

**Figure 1.** Growth process of a single crystal of MAPbBr$_3$ on a VACNTs forest. (a) Photographs of growing MAPbBr$_3$ crystals acquired at different time intervals. (b) Schematic representation of the growth of MAPbBr$_3$ single crystals on a VACNTs forest. (c) Image of VACNTs grown on a silicon substrate, (d) Image of a single crystal of MAPbBr$_3$ grown on top of the VACNT forest, and (e) Image of a detached sample showing the 'crystal – CNTs' junction. (f) Low- , (g) medium- and (h) high-magnified scanning electron microscopy images of the junction between the VACNTs and the MAPbBr$_3$ single crystal.

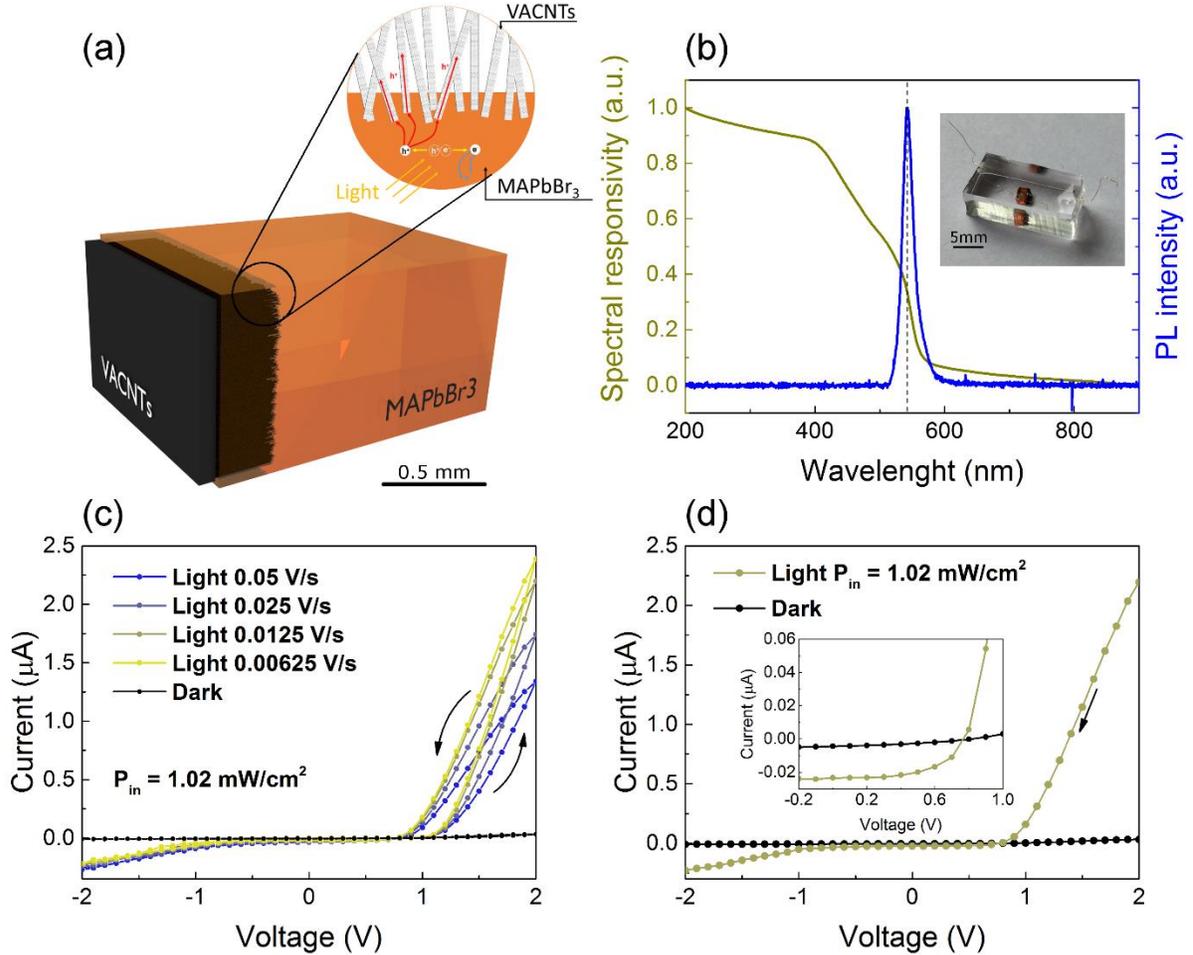

**Figure 2.** (a) Schematic illustration of the sample and the interface between MAPbBr3 and VACNTs, highlighting the photo induced charge transfer mechanism under reverse or zero bias voltage. (b) Spectral responsivity of the MAPbBr$_3$ single crystal-based photo-detector: photoluminescence (PL) spectrum excited with $\lambda_{exc}$ = 470 nm (blue trace) and spectral evolution of the photocurrent (golden trace). (inset) Image of the PDMS-embedded detector. (c) *I-V* measurements of the embedded device in dark and under visible light (intensity 1.02 mW/cm$^2$), for different scan rates. The voltage was swiped from 0 V to +2 V/-2 V and back. When increasing the voltage, the current is lower than while decreasing it. (d) *I-V* measurements from 2 V to -2 V with a scanning speed of 0.0125 V/s. (inset) Magnified view of the region around zero bias voltage.

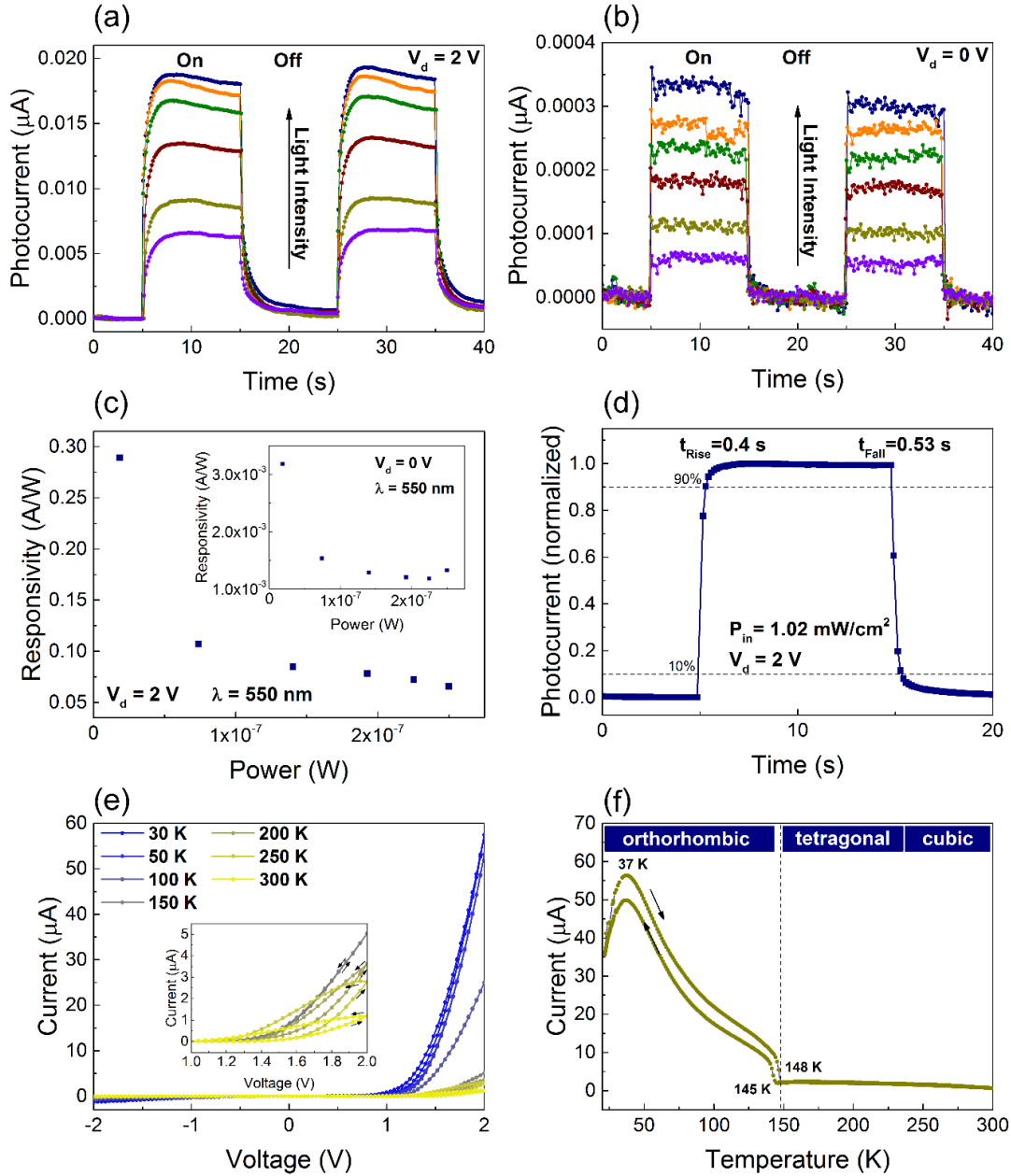

**Figure 3.** (a) Transient photoresponse on-off characteristics of the photodetector under illumination with 550 nm light at different light intensities (18 nW, 74 nW, 149 nW, 193 nW, 225 nW and 250 nW), at bias voltage of 2 V, after encapsulation and (b) at bias voltage of 0 V, before encapsulation. (c) Responsivity of the perovskite photodetector at bias voltage of 2 V (main plot) and at bias voltage of 0 V (inset). (d) Single normalized photocurrent response on-off cycle of the photodetector at bias voltage of 2 V, under visible light of intensity 1.02 mW/cm$^2$. (e) The temperature evolution of the *I-V* plots collected for the photodetector under visible light (intensity 0.77 mW/cm$^2$). (inset) Magnification of the lower current region for *I-V* plots at higher temperature. (f) The temperature dependence of the photocurrent at bias voltage of 2 V under illumination with visible light (0.586 mW/cm$^2$).